\documentclass[conference, a4paper]{IEEEtran}
\IEEEoverridecommandlockouts
\usepackage{cite}
\usepackage{amsmath,amssymb,amsfonts}
\usepackage{algorithmic}
\usepackage{graphicx}
\usepackage{textcomp}
\usepackage{xcolor}
\def\BibTeX{{\rm B\kern-.05em{\sc i\kern-.025em b}\kern-.08em
    T\kern-.1667em\lower.7ex\hbox{E}\kern-.125emX}}

    \def\BibTeX{{\rm B\kern-.05em{\sc i\kern-.025em b}\kern-.08em
    T\kern-.1667em\lower.7ex\hbox{E}\kern-.125emX}}
    
\begin{document}

\title{Electric Vehicle Charging Profile Forecasting Using Hybrid Models
\\
\thanks{This study was partially carried out within the project "EVFLEX: soluzioni per la ﬂessibilità dei veicoli elettrici nella rete" and received funding from MASE - Ministero dell’Ambiente e della Sicurezza Energetica (MISSION INNOVATION 2.0, MI\_FAE\_00360). This manuscript reflects only the authors’ views and opinions.}
}

\author{\IEEEauthorblockN{Riccardo Ramaschi}
\IEEEauthorblockA{\textit{Department of Energy} \\
\textit{Politecnico di Milano}\\
Milan, Italy \\
riccardo.ramaschi@polimi.it}
\and
\IEEEauthorblockN{Mario Paolone}
\IEEEauthorblockA{\textit{Distributed Electrical Systems Laboratory} \\
\textit{École Polytechnique Fédérale de Lausanne}\\
Lausanne, Switzerland \\
mario.paolone@epfl.ch}
\and
\IEEEauthorblockN{Sonia Leva}
\IEEEauthorblockA{\textit{Department of Energy} \\
\textit{Politecnico di Milano}\\
Milan, Italy \\
sonia.leva@polimi.it}
}

\maketitle

\begin{abstract}
Electric Vehicle (EV) fast charging stations require forecasting techniques both at the single charger level and aggregated level. While for the latter several models exist, forecasting individual EV charging profiles is still underexplored in literature. However, such methods may be potentially used by battery-aware scheduling, leading to a more granular update of the charging station aggregated forecast and provide a more accurate estimation of EVs departure times. Nonetheless, the variable extent of available information in time and in different settings could jeopardize these benefits. For this reason, we propose a hybrid and lightweight method to estimate the EV charging profile before and during the charging process. Besides evaluating this method on multiple EVs from a public dataset, we also assess the impact of different level of information in the time transposition of the charging profile.
\end{abstract}


\section{Introduction}
Global Electric Vehicle (EV) penetration is constantly increasing, recording a 25\% additional sales in 2024 with respect to 2023 \cite{EVoutlook25}. This growth is supported by the parallel expansion of the charging infrastructure, which experienced a 28\% year-on-year increase in charging points in 2025 \cite{EAFO} in the European Union, most of which are part of DC Fast Charging Stations (FCS). This kind of infrastructure, that experienced approximately a 50\% increase in 2025, is mainly equipped with charging points with at least 50 kW charging rate.

Since the aggregated effect of EV demand in FCS is considered of utmost importance for power distribution grids' energy \cite{Energy} and congestion \cite{Congestion} management, the majority of literature focus on the aggregated power demand. In fact, researchers often model the EV demand as a single time series that is fed into statistical methods leveraging historical endogenous and exogenous data \cite{aggregate}. On the contrary, forecasting each individual EV charging profile is underexplored in literature, despite the potential benefits of such information. In fact, predicting the EV charging profile may be used by battery-aware power scheduling \cite{simolin_1}, to provide a more granular update of the charging station aggregated forecast \cite{Alfredo} and to estimate the individual EVs departure times \cite{Time}.

Despite some works exist on battery charging profile prediction, few focus on the EV domain and on its unique characteristics and constraints, i.e. the lack of granular and public datasets \cite{review_dataset} and the dependency of the charging profile to the proprietary battery management system \cite{review_charging}. Among these, most are limited by private datasets \cite{psoc}, a small set of different EV manufacturer \cite{Time} and power-constrained charging sessions, where the maximum charging rate is limited by the capabilities of the charging station \cite{Alfredo, simolin_1}. To the best of the author knowledge, two works address the charging profile estimation within FCS focusing on the EV required power, thus avoiding power-constrained charging sessions. Li et al. \cite{psoc} proposed a deep learning framework that initiates the charging profile estimation and further refines it throughout the charging session. Besides relying on a private dataset, the work represents a strong contribution; however, the model employs architectures that could be computationally demanding if retraining are needed over a rolling horizon. Simolin et al. \cite{epfl_db_analysis} modeled charging profiles of heterogeneous EVs, from a public dataset \cite{desl_db} of a Level 3 (L3) FCS, through a simple polynomial model that still overcame existing baselines.

In this work, we present and assess the performance of a comprehensive charging profile forecasting tool and we also plan to fill a gap in the charging profile time transposition. This consists in the transformation of the EV power vs State-of-Charge (SoC) into the power vs time curve. In fact, while both \cite{psoc} and \cite{epfl_db_analysis} considers the information needed to perform this transposition known, this is not always the case and it may impact the process \cite{simolin_3}. Therefore, we also identify the impact of the information stream between the EV and the charging station in the power vs time curve definition, e.g. arrival/departure SoC, arrival/departure time and battery capacity. Our contribution is threefold. First, we outline the forecasting tool. Second, we quantify the forecasting accuracy of this model. Third, we perform an ablation-like analysis on the availability of the abovementioned features.


\section{Methodology} \label{meth}
The individual EV charging profile forecasting tool is made of two main instances. The first one produces an initial guess on the charging profile prior to connection (Section \ref{before}). At this step, we assume known the arrival time $t_a$, either through a booking system or through real-time sensing. The second one provides a continuous refinement after connection (Section \ref{after}). At this step, considering a L3 FCS \cite{ISO15118-20}, the infrastructure has access to the battery capacity, the realized charging profile $\mathcal{H}_{t_i} = \{ (SoC_t, P_t) \}_{t=t_a}^{t_{i-1}}$, the user-defined departure SoC $SoC_{t_d}$, while the departure time $t_d$ is inferred. The overall model and its time transposition are described in Section \ref{time}.

\begin{figure}[t]
    \centering
    \includegraphics[width=0.9\linewidth]{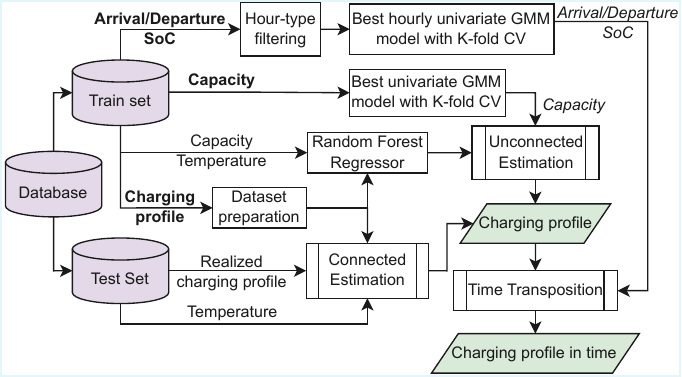}
    \caption{Overall model for charging profile estimation and its time transposition. In bold the target of the different models, in italic the potentially unknown information.}
    \label{fig:intra_ev}
\end{figure}

\subsection{Unconnected EV} \label{before}
Following the indications of \cite{temperatures_1} on ambient temperature impact at EV arrival time on charging profiles, we include $T_{t_a}$ as a feature in our unconnected forecasting model, together with the EV battery capacity that we indicate as $C$. The objective is to estimate the charging profile, defined as the mapping between the SoC and the charging power over the full range $\text{SoC} \in [0,100]\%$, with a 1\% resolution, that we define as $\mathcal{CP}_{t_a} = \{ P(SoC)\}_{SoC=0\%}^{100\%}$. We first construct a database of historical sessions through SoC-based interpolation of the charging profiles. Since not all sessions span between 0 and 100\%, we handle missing data using constant extrapolation at the boundaries. That is, we extend the profile using its first and last available values, if needed. This results in a fixed-length representation of the training dataset. We then formulate the forecasting task as a supervised learning problem, employing a Random Forest (RF) regressor targeting $\mathcal{CP}_{t_a}$.

\subsubsection{Capacity estimation}
$C$ could be either the actual value, if promptly communicated, or an estimation. Nonetheless, we generalize by proposing a capacity estimation through the Gaussian Mixture Model (GMM), a probabilistic and parametric method used for clustering and forecast. Inspired by the work in \cite{Grid_aware}, where GMM is used as an EV forecaster over the same dataset \cite{desl_db}, we employ an univariate GMM with K-fold Cross Validation (CV) \cite{kFold} to identify the most suitable model. While in testing phase, the identified model generates a random sample from the learned probability distribution.

\subsection{Connected EV} \label{after}
Once connected, we build a rolling-horizon framework to update the charging profile with a time granularity $\Delta t$, so that we continuously refine the unconnected first guess and the following estimations. For this purpose, to exploit the historical realizations of other EVs and with the goal of developing a lightweight solution suitable for rolling-horizon application, we propose a trivial statistical method based on euclidean distances. Let us consider an EV with a battery capacity $C$ arriving at time $t_a$ and requiring a power $P_{t_a}$ at $SoC_{t_a}$ during the first time step. At each timestep $t_i$, at $SoC_{t_i}$, our model aims at finding $\mathcal{CP}_{t_i} = \{ P(SoC)\}_{SoC=SoC_{t_i}}^{SoC_{t_d}}$. Therefore, at time $t_i$ we create an input $x_{t_i}=[C, P(SoC_{t_a}), ..., P(SoC_{t_i-1})]$ that is compared with a sliced historical realizations matrix:
\[
\mathbf{X}_{t_i} =
\begin{bmatrix}
C^{(1)} & P^{(1)}(\mathrm{SoC}_{t_a}) & \cdots & P^{(1)}(\mathrm{SoC}_{t_{i-1}}) \\
C^{(2)} & P^{(2)}(\mathrm{SoC}_{t_a}) & \cdots & P^{(2)}(\mathrm{SoC}_{t_{i-1}}) \\
\vdots & \vdots & \ddots & \vdots \\
C^{(N)} & P^{(N)}(\mathrm{SoC}_{t_a}) & \cdots & P^{(N)}(\mathrm{SoC}_{t_{i-1}})
\end{bmatrix}
\]
where each line is indicated as $x_{t_i}^{(j)}$. Here we apply the euclidean distance to find the closest $x_{t_i}^{(j^*)}$ and we update the charging profile accordingly for each time $t_i$:
\begin{align*} 
    &j^* = \arg\min_{j} \left\| x_{t_i} - x_{t_i}^{(j)} \right\|_2\\
&\mathcal{CP}_{t_i}=\{ P^{(j^*)}(SoC)\}_{SoC=SoC_{t_i}}^{SoC_{t_d}}
\end{align*}

\subsection{Overall model and its time transposition} \label{time}
During the whole charging session, the model generates $\mathcal{CP}_{t_i}$ for each $t_i\in[t_a; t_d]$. Before connection, information on arrival/departure SoC and departure time are potentially unknown as the battery capacity. For the latter, we developed a GMM-based estimation described in Section \ref{before}. Since all the relevant information is known once the connection occurs, we focus on the impact of these parameters prior to connection. We develop a model on SoC estimation (both for arrival and departure SoC) rather than $t_d$, that can be inferred once known (or estimated) the SoCs. We employ a GMM-based framework also for these variables' forecasting. In particular, we compute for both variables the hourly best (through K-fold CV) univariate GMM model, based on the principle that the arrival and departure SoC are mainly influenced by the corresponding time. Then, in the testing phase, the hourly models are chosen according to the arrival hour and the departure hour (that we assume known in a FCS) and generate a random sample from the learned probability distribution for both variables. The time transposition is then performed, according to the available information, through a simple EV surrogate model. We show in Figure \ref{fig:intra_ev} the overall model.

\section{Case Study} \label{data}
We present the dataset in Section \ref{dataset}, the experiments in Section \ref{cases} and the hyperparameter tuning in Section \ref{hyperparam}.

\begin{table}[t]
\centering
\caption{Hyperparameter fine-tuning for the RF model}
\label{tab:rf_hyperparameters}
\begin{tabular}{lcc}
\hline
\textbf{Hyperparameter} & \textbf{Search Space} & \textbf{Optimal Value} \\
\hline
Estimators & $[100, 1000]$ (step 50) & 600 \\
Max depth & $[2, 50]$ & 47 \\
Min samples split & $[2, 20]$ & 2 \\
Min samples leaves & $[1, 20]$ & 18 \\
Max features & \{sqrt, log2\} & sqrt \\
Bootstrap & \{True, False\} & True \\
Criterion & \{Squared, Absolute\} & Squared \\
\hline
\end{tabular}
\end{table}

\subsection{Dataset} \label{dataset}
For this analysis, we use the public dataset from the Distributed Energy Sources Laboratory (DESL) \cite{desl_db} of the École Polytechnique Fédérale de Lausanne, Switzerland. This dataset, collecting charging sessions measurement from two different charger, comprises 1878 L3 charging sessions from April 2022 to July 2023. They include 64277 measurements with a 1-minute resolution ($\Delta t$). We focus on the time-stamped recorded SoC, the required power, the battery capacity and the control flag, identifying those session that have been controlled. We then include with Open-Meteo API \cite{OpenMeteo2023} the historical recorded ambient temperature at the DESL location. Further information on the database can be found at \cite{desl_db}, while a complete dataset description is presented in \cite{epfl_db_analysis}. We split the dataset into training and test sets using a 70/30\% split.

\subsection{Experiments} \label{cases}
We carry out two sets of experiments. The first, i.e. the charging profile forecasting experiment, employs the overall dataset. The second, i.e. the time transposition experiments, only exploits those charging process that were not controlled (power provided equal to power required), that are 162 in the test set. This is because we need to isolate the intrinsic charging profiles of electric vehicles from external control actions. For this reason, on this second analysis, we focus on available information before the charging process, i.e. $SoC_{t_a}$, $SoC_{t_d}$ and $C$, and their impact into the time transposition of the initial charging profile forecast (unconnected). These information can be known or estimated, therefore we compare eight scenarios: i) perfect information, all variables are available, ii)-vii) all intermediate combinations, and viii) no information, all variables are estimated.

\subsection{Hyperparameters} \label{hyperparam}
Both the capacity and SoC GMM-based tools use a K-fold CV to identify the best number of components $MC$, from 1 to the maximum number of mixture components $MC_{max}$. $MC$ is identified by minimizing the cross-validated minimum absolute error. Following the indication in \cite{Grid_aware}, we set $MC_{max}=11$ and $K=5$. For what concerns the RF regressor, we carried out a hyperparameter tuning through Optuna \cite{optuna}. We set a wide search space on several hyperparameters, while at each trial (100 trials as a trade-off between accuracy and computational complexity) Optuna samples a candidate configuration using a Tree-structured Parzen Estimator sampler. We show in Table \ref{tab:rf_hyperparameters} the hyperparameters' search space and optimal value. We evaluate the RF regressor again with K-fold CV ($K=5$), employing pruning mechanism to terminate unpromising configurations early. 

Both GMM models and the RF regressor are implemented through the Scikit-Learn Python package \cite{10.5555/1953048.2078195}.

\begin{figure}[t]
    \centering
    \includegraphics[width=\linewidth, height=0.47\linewidth]{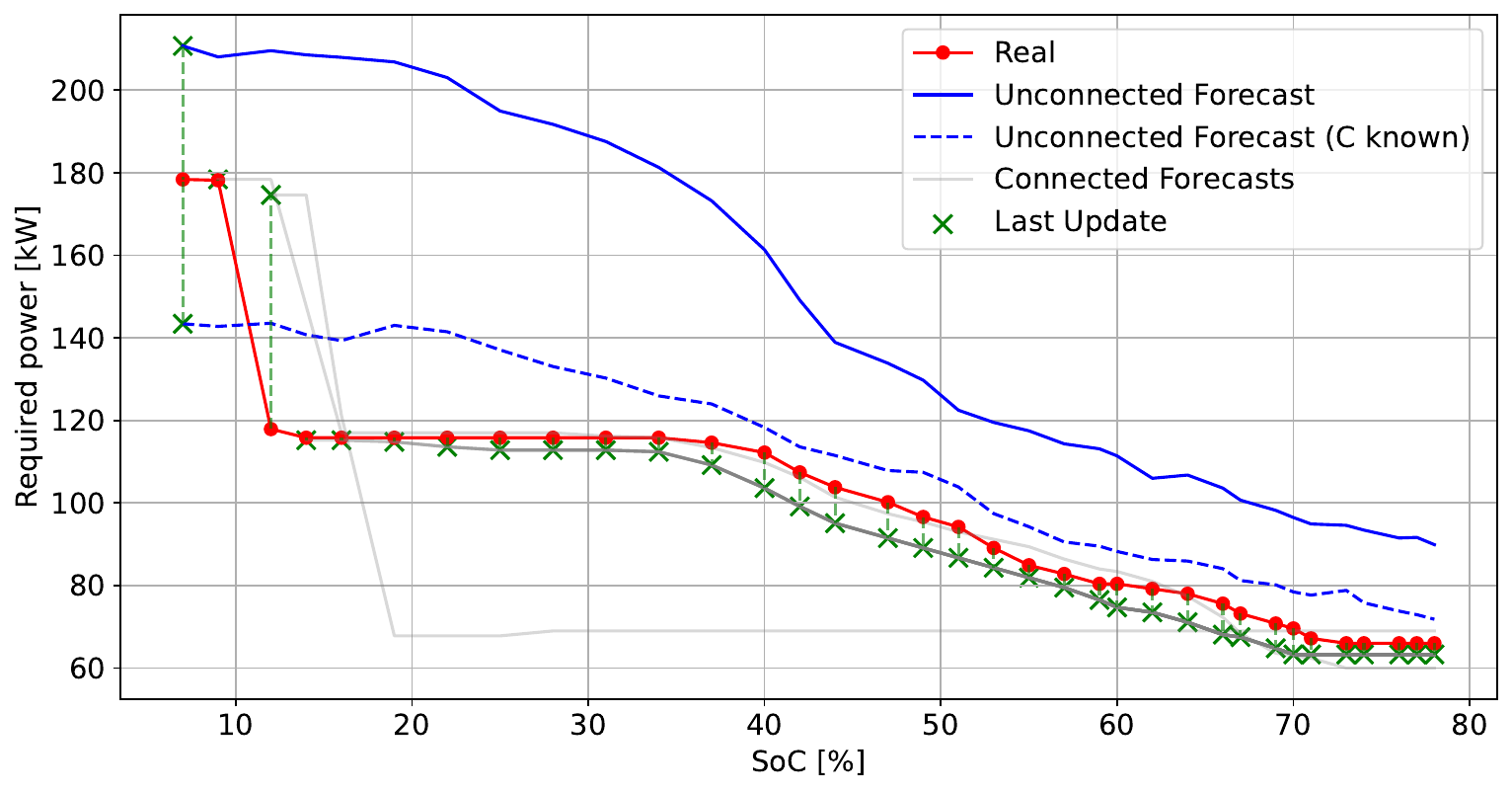}
    \caption{Overall model functioning example, where $C$ was estimated at 73.9 kWh while actual value was 57.4 kWh. SoC mismatch is not considered here.}
    \label{fig:example}
\end{figure}

\begin{figure}[t]
    \centering
    \includegraphics[width=0.96\linewidth]{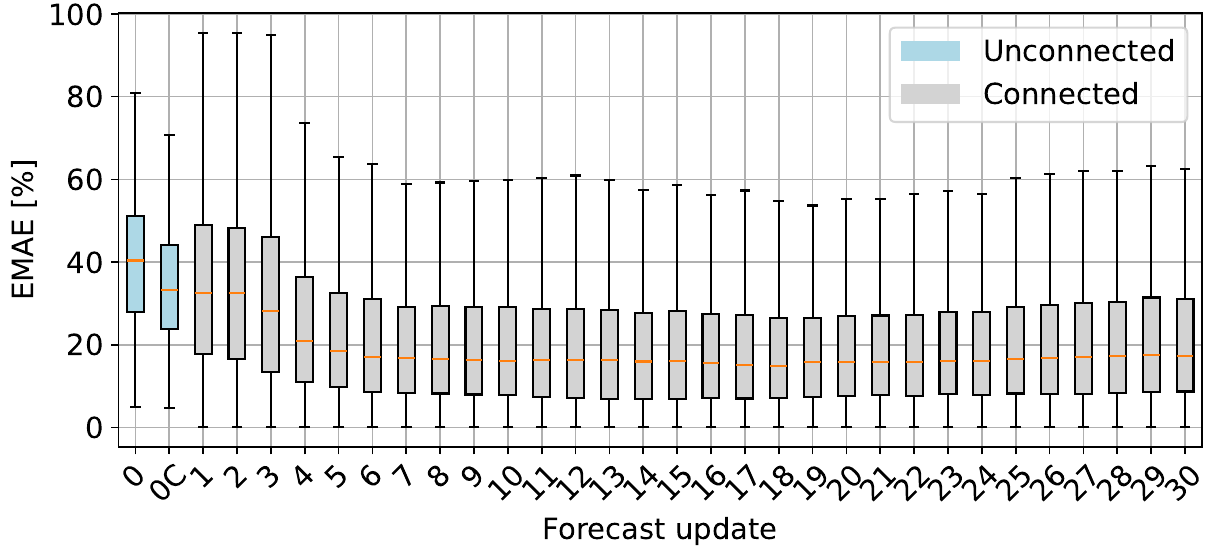}
    \caption{EMAE distribution trend at different forecast update.}
    \label{fig:emae}
\end{figure}

\section{Results} \label{results}

\subsection{Charging profile forecasting} \label{psoc}
We start by illustrating an example of the overall method functioning in Figure \ref{fig:example}. First, the unconnected single EV forecast is created (blue): in this case, we report both the output of the RF estimation when the battery capacity is known (dashed) and estimated (solid). As expected, the former leads to more accurate estimation. Second, we perform for each minute an update of the charging profile according to the connected forecast instance. This continuous update generates several predicted charging profiles (in grey), whose first estimation is shown in green. As more realization occur, the forecast is refined before reaching a stable forecast, that is particularly accurate in this example.

To quantify the improvement of this updating forecast, we present in Figure \ref{fig:emae} the distribution of the Envelope-weighted  Mean Absolute Error (EMAE) \cite{EMAE} over the entire test set throughout the charging process. For statistical relevance, we show the first 30 iterations in addition to the two unconnected models (one estimating the battery capacity, defined as $0$, and one knowing it, defined as $0C$). Starting from the unconnected forecasts, this figure quantifies the effect of knowing the capacity before the connection, with a reduction of about 10\% in the median EMAE. As the forecast updates, after an initial plateau in the first two time steps, the overall distribution shifts downward between the third and the tenth update. Afterwards, the median EMAE remains stable around 16\% until the twentieth update, where it starts slowly to increase again (together with the overall distribution). This result, that can be motivated by the purely data-driven nature of the connected forecast, suggests to avoid to continuously refine the charging profile forecast, but rather to stop after 10 iterations.

\begin{figure}[t]
    \centering
    \includegraphics[width=0.85\linewidth]{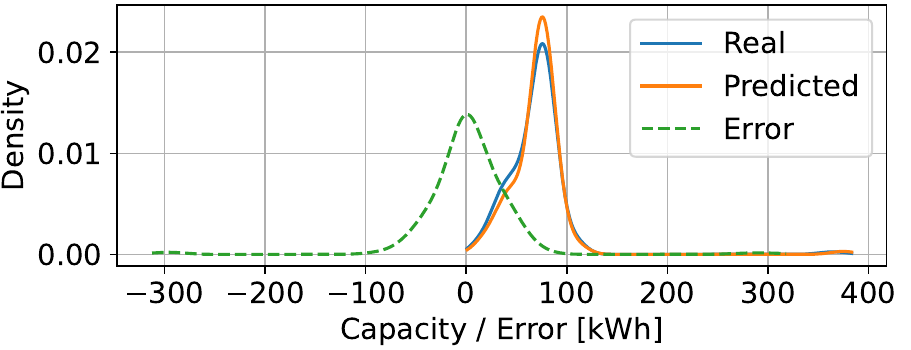}
    \caption{Capacity estimation model performance.}
    \label{fig:capacity}
\end{figure}

\begin{figure}[t]
    \centering
    \includegraphics[width=\linewidth]{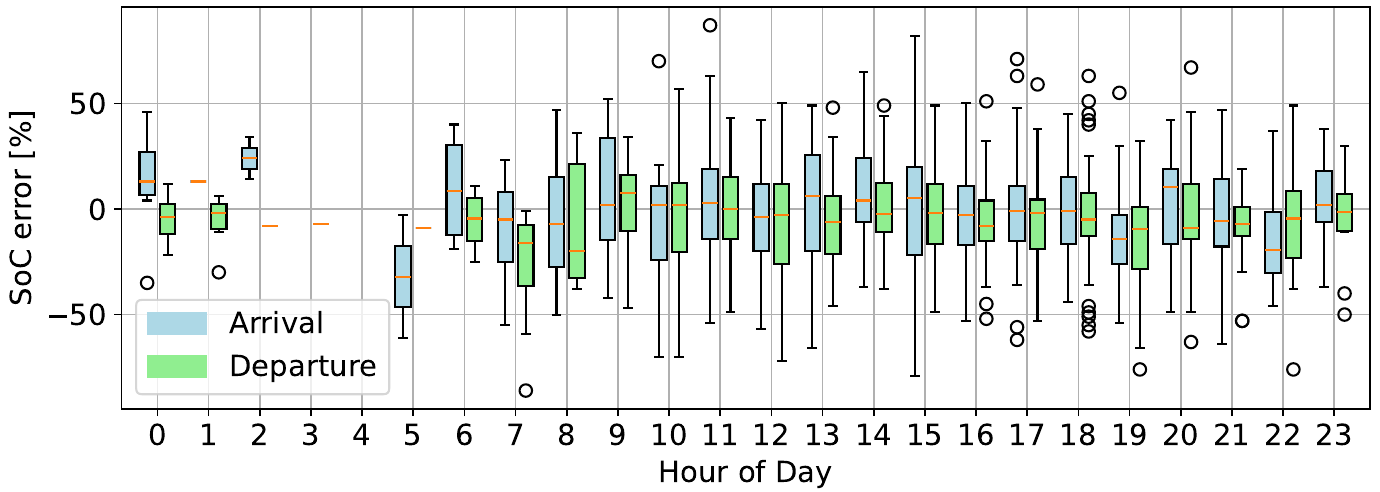}
    \caption{Hourly distribution of $SoC_{t_a}$ and $SoC_{t_d}$ errors.}
    \label{fig:soc}
\end{figure}

\begin{figure}[t]
    \centering
    \includegraphics[width=\linewidth, height=0.5\linewidth]{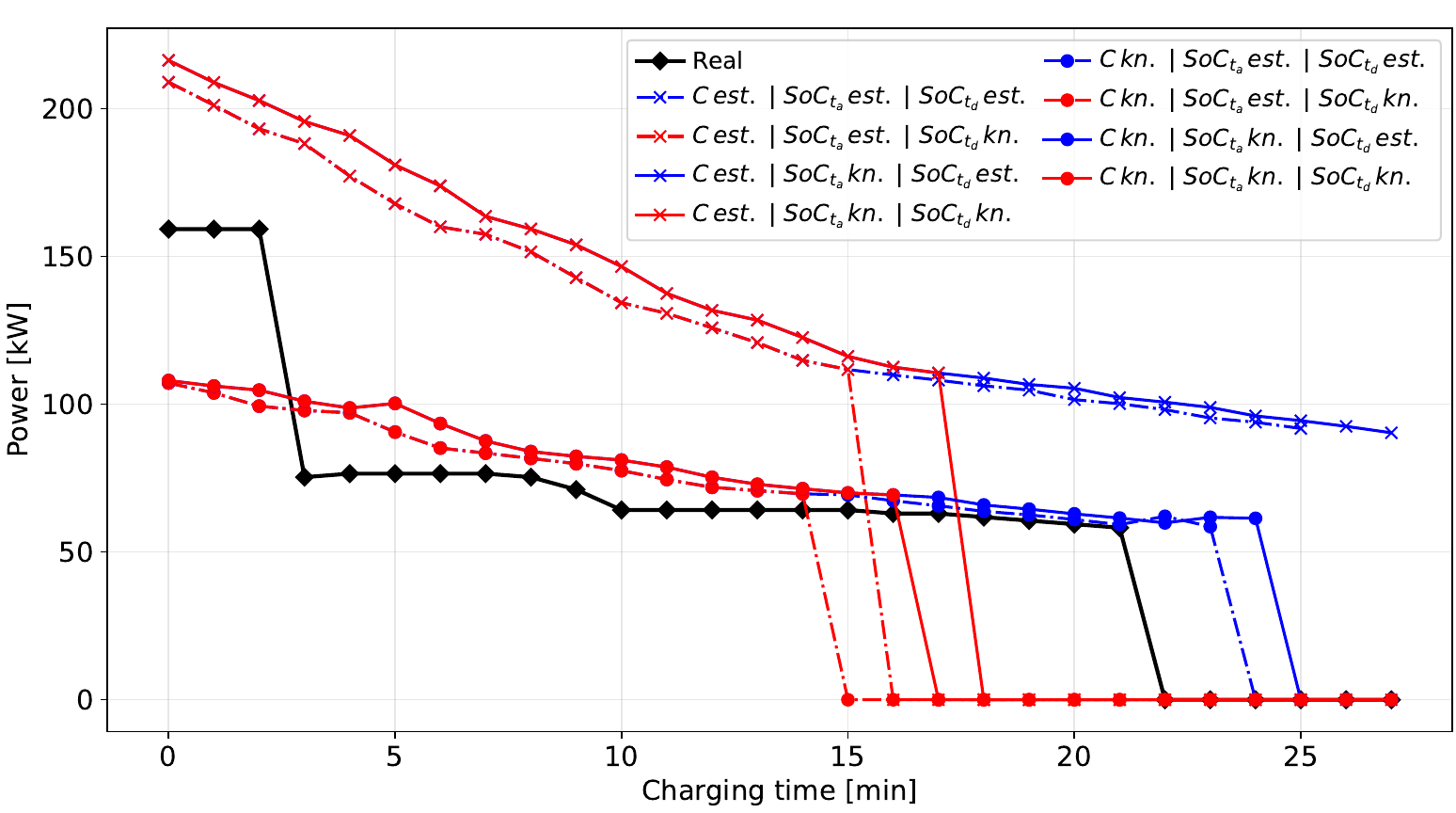}
    \caption{Time transposition example over the eight identified scenarios: marker identifies $C$ (est. x, kn. o), color $SoC_{t_d}$ (est. blue, kn. red) and style $SoC_{t_a}$ (est. dashed, kn. solid).}
    \label{fig:example_time}
\end{figure}

\begin{figure}[t]
    \centering
    \includegraphics[width=\linewidth, height=0.75\linewidth]{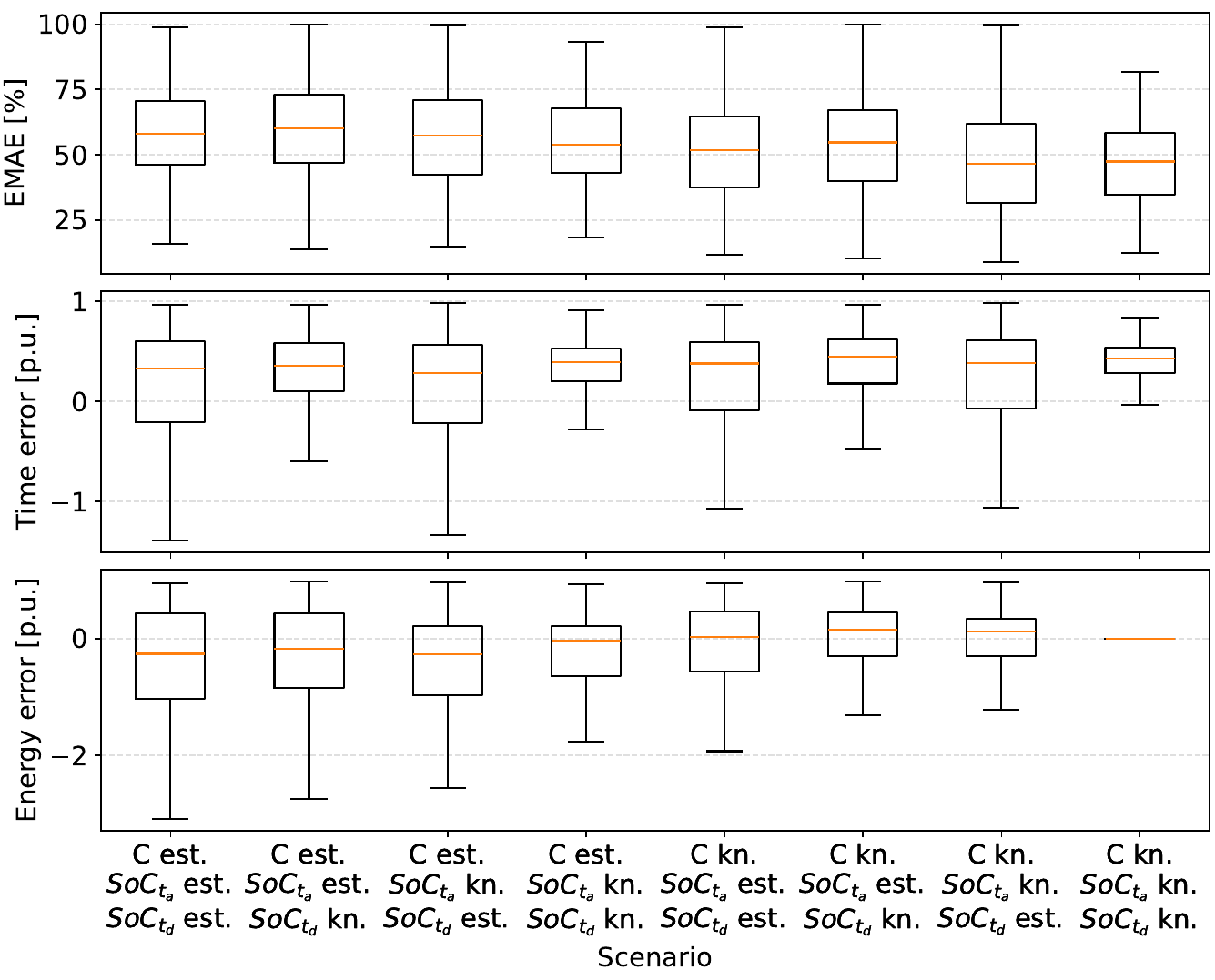}
    \caption{Scenario evaluation impact on three accuracy metrics.}
    \label{fig:boxplot}
\end{figure}

Finally, we show in Figure \ref{fig:capacity} the Kernel Density Estimates (KDE) of real and predicted battery capacities, together with the corresponding prediction error distribution, to assess the agreement between model outputs and ground truth. The figure shows how the GMM-based model successfully replicates the distribution, despite the wide error distribution range that may jeopardize the initial forecast.

\subsection{Time transposition of the unconnected estimation} \label{ptime}

Before analyzing the 162 uncontrolled charging sessions identified in Section \ref{cases}, we show in Figure \ref{fig:soc} the hourly distribution of the arrival and departure SoC error (real minus predicted). The figure shows a slight tendency of overestimation of both SoCs, as typically the median lies in the negative part of the plot. This bias is more pronounced for departure SoC. Relying on a limited amount of data both in training and in testing (e.g., no data between 2 and 5 a.m.), the hourly SoC models are not as accurate as the battery capacity model (see Figure \ref{fig:capacity}) to replicate the real distribution. Nonetheless, the absolute medians for both errors in each hour seldom overcome 25\%, being typically around 10\%.

Since the key variables become available upon EV connection, the time transposition analysis focuses on mapping the unconnected profile estimation into the time domain. We show an example of the time transposition across eight scenarios in Figure \ref{fig:example_time} (estimated is shortened to est. and known to kn.), where $C$ was estimated at 79.5 kWh, $SoC_{t_a}$ at 22 \% and $SoC_{t_d}$ at 81 \%, with the real values at 41.5 kWh, 17\% and 64\%, respectively. According to the capacity estimation error, also the corresponding charging profiles estimation differ significantly (as well as the time transposition shape), with the power curve with $C$ known being particularly accurate. On the other hand, errors in $SoC_{t_a}$ and $SoC_{t_d}$ mainly affect the temporal alignment of the power curve, leading to shifts in the charging horizon. Since the effects of these three variables accumulate, the error propagation is not trivial to analyze. While capacity alone influences the charging profile forecast, all the three variables might lead to significantly different power versus time curves, as well as different estimated charging time and scheduled energy.

For this reason, we show in Figure \ref{fig:boxplot} the distribution on each scenario for what concerns three accuracy metrics, i.e. charging profile, charging time and scheduled energy. In the upper graph, the charging profile accuracy is visualized through the EMAE of the power curves (as Figure \ref{fig:example_time}). As discussed above, all variables strongly affect the profile transposition, thus leading to higher EMAE distributions with respect to those shown in Figure \ref{fig:emae}. When all the variables are known, the distribution is shifted downward and the median lies around 45\% together with the scenario with $SoC_{t_d}$ estimated (that still exhibits a wider distribution), while all the other scenarios show a median above 50\%. In general, errors on $C$ lead to systematic errors across scenarios since the system operates with a wrong scaling of available energy. In contrast, errors on SoC have a different impact: $SoC_{t_a}$ leads to consistent EMAE median decrease when known, while $SoC_{t_d}$, counterintuitively, leads to equal or higher median EMAE when known, probably due to the bias identified in Figure \ref{fig:soc}.

In the middle and the lower graph, we show the per unit estimated charging time error and scheduled energy error (positive means underestimation, negative means overestimation). Starting from the time error, scenarios impact the distribution rather than the median, that remains around 0.3 across different combinations, meaning that the model is biased towards underestimation in charging time. In this case, knowing $C$ leads to fewer and less critical overestimation while the remaining distributions exhibit similar behaviors. Due to the identified bias in Figure \ref{fig:soc}, the impact of knowing $SoC_{t_d}$ is to tighten the distribution in the underestimation quadrant. On the other hand, $SoC_{t_a}$ has negligible effect on this metric. For what concerns the energy error, we do not considered the perfect information scenario as scheduled energy is inherently known. When no information is available, the distribution covers a wide range of errors, that is reduced by having access to any variable. In fact, with respect to the no information scenario, any other scenarios exhibit narrower distributions with medians closer to 0. In particular, information on $C$ make the distributions more symmetric while inverting the bias from overestimation to underestimation, whereas information on $SoC_{t_a}$ and $SoC_{t_d}$ only reduces the distributions spread. 

Overall, when transposing the unconnected estimation of the charging profile in the time domain, capacity governs bias, while SoC information mainly affect dispersion.

\section{Conclusions} \label{concl}
This work assessed the performance of a charging profile forecasting tool using a hybrid and lightweight model for unconnected and connected updates on a public dataset. First, we employ a RF regressor to predict the charging profile. Then, we envision a rolling-horizon statistical update based on euclidean distances. We showed how this sequential refinement improves the initial estimation up to the tenth iteration, reducing the median EMAE from 40\% to 16\%. With the goal of evaluating the impact of different information availability on the charging profile initial estimation transposition into the time domain, we proposed GMM-based models for battery capacity and for arrival/departure SoC prediction. Evaluating eight scenarios on charging profile, charging time and scheduled energy accuracy, the results showed that the errors compound, but capacity has more impact on the bias while the SoCs on the spread of those metrics' distributions. As a future work, we plan to enhance the proposed methodology on two directions. First, integrating a wider database, and second, considering different ML implementation especially for the connected instance. Finally, we will benchmark this future model with other state-of-the-art baselines.

\bibliographystyle{IEEEtran}
\bibliography{literature}

\end{document}